\documentclass[doublecol]{epl2}

\usepackage{SIunits}
\usepackage{xspace}
\usepackage{epsfig}
\usepackage{wrapfig}
\usepackage{subcaption}
\usepackage{verbatim}
\usepackage{ifthen}
\usepackage{amsmath}
\usepackage[none]{hyphenat}

\begin{document}
% @1 +_defs

\newcommand{\Mark}{%
\par
\vspace{5mm}
\noindent------------------------------------------------------
\par
}

% ../2013/_defs.tex
% app/_defs.tex

\newcommand{\ie}{i.e.{}\xspace}			% id est
\newcommand{\versus}{\emph{vs}\@\xspace}
\newcommand{\dc}{d.c.\xspace}
\newcommand{\viceversa}{\emph{vice versa}\xspace}
\newcommand{\cf}{cf.\@\xspace}
\newcommand{\eg}{e.g.\@\xspace}
\newcommand{\enroute}{\emph{en route}\xspace}
\newcommand{\ansatz}{\emph{ansatz}\xspace}
\newcommand{\viz}{viz\xspace}

\newcommand{\AU}{\mathrm{AU}}

\newcommand{\Vector}[1]{\ensuremath{\mathbf{#1}}}
\newcommand{\Exp}[1]{\left\langle #1 \right\rangle}

\newboolean{notes}
\setboolean{notes}{false}

\ifthenelse{\boolean{notes}}{
	\newcommand{\Note}[1]{\marginpar{\small\it\textcolor{BrickRed}{#1}}}
	\newcommand{\PatNote}[1]{\marginpar{\small\it\textcolor{Emerald}{#1}}}
	\newcommand{\FigNote}[1]{\marginpar{\small\it\textcolor{Green}{#1}}}
	\newcommand{\InFigNote}[1]{\marginnote{\small\it\textcolor{OliveGreen}{#1}}}
	\newcommand{\EqNote}[1]{\marginpar{\small\it\textcolor{Blue}{#1}}}
	\newcommand{\SecNote}[1]{\label{ss:#1}\marginpar{\small\it\textcolor{Purple}{#1}}}
}{
	\newcommand{\Note}[1]{}
	\newcommand{\PatNote}[1]{}
	\newcommand{\FigNote}[1]{}
	\newcommand{\InFigNote}[1]{}
	\newcommand{\EqNote}[1]{}
	\newcommand{\SecNote}[1]{\label{ss:#1}}
}

% @1 +main
% ------------------------------------------------------------------
% @2 +title
\title{Observational evidence for travelling wave modes bearing distance proportional shifts}
\shorttitle{Travelling wave modes bearing distance proportional shifts}

\author{V.~Guruprasad}
\shortauthor{\it{prasad}}

%\email{prasad@inspiredresearch.com}
\institute{Inspired Research, New York 10509, USA.}

%\pacs{95.55.Pe}{Lunar, planetary, and deep-space probes}
%\pacs{04.80.?y}{Experimental studies of gravity}

\renewcommand{\baselinestretch}{1}
\abstract{
% @2 +abs
Discrepancies of range between
	the Space Surveillance Network radars
and
	the Deep Space Network
in tracking
	the 1998 earth flyby of NEAR,
and between
	ESA's Doppler and range data in
		Rosetta's 2009 flyby,
reveal
	a consistent excess delay, or lag,
equal to
	instantaneous one-way travel time
		in the telemetry signals.
These lags readily explain
	all details of the flyby anomaly,
and are shown to be symptoms of
	chirp d'Alembertian travelling wave solutions,
relating to
	traditional sinusoidal waves
by a rotation of
	the spectral decomposition
due to the clock acceleration caused by
	the Doppler rates during the flybys.
The lags thus relate to
	special relativity,
but yield
	distance proportional shifts
		like those of cosmology
	at short range.}

\maketitle
% @2 +body

% background {{{

The fourth-power power law limits direct radar tracking,
as provided by
	the Space Surveillance Network (SSN),
to about
	the range of geostationary orbits
	($36,000~\kilo\metre$).
For tracking spacecraft on deep space missions,
NASA's Deep Space Network (DSN) uses 
	the telemetry signal
returned by
	the phase-coherent onboard transponder
for both
	range and Doppler measurements,
using
	modulated range codes
and
	the carrier,
respectively,
as detailed in
	\cite[\S{III}]{Anderson2002a}.
Using spin-stabilized spacecraft,
this approach achieves
	sufficient precision for
		tests of general relativity
	\cite{Bender1989,Vincent1990,Anderson1993}.
Over decades,
this approach has led to
	four space anomalies
	\cite{Anderson2009},
of which the best known,
	the Pioneer anomaly,
has now been traced to
	an overlooked radiation reaction
	\cite{Turyshev2012}.

The present work fully explains
	the earth flyby anomaly,
without assuming
	dark matter
	(\cf \cite{Adler2011}),
or
	modifications to gravitation theory
	(\cf \cite{Anderson1998,Anderson2008}).
A broader result is
	a local mechanism
that relates more closely to
	special relativity and propagation,
yet yields
	distance proportional spectral shifts
		along with time dilations,
which are thought to need
	an expanding space-time
	(\cf \cite{SandageLubin2001,Wolf1987a,Wolf1987b,Wolf1989}).

The distance proportionality is given by
	large negative residuals of
		the SSN data
	\cite{Antreasian1998},
against
	the DSN-estimated trajectory,
which,
	barring contrived hypotheses,
can only mean either that
	the SSN radar echoes were superluminal
		specially during the flybys,
or that
	the DSN Doppler and range data had
		an excess delay.
These residuals have been omitted in
	later discussions
	\cite{Anderson2006,Anderson2008,Gerrard2008,Mbelek2009,
		McCulloch2008,Turyshev2010,Nyambuya2010,
		Bertolami2011,Paramos2012,Lorio2013},
as they exceed
	the SSN resolutions,
but radar cannot have less than
	two-way delay
or 
	large variations
regardless of
	processing errors.

The excess delay equals light time
	for the instantaneous range,
and
the residuals match the radial distance that
	the spacecraft would travel
		in that time.
Corresponding shifts
	in the telemetry spectra
are implied by the consistency of
	the demodulated range codes with
the delay in the carrier
	affecting the DSN Doppler.
Both effects are traced to
	large radial Doppler rates
		not seen with orbiting satellites;
their general absence beyond orbit range
	is also explained below by
the spectral selection
	in the receiving process.

The core contribution,
	with the broadest significance,
is the explanation of
	the delays and the shifts themselves
as properties of
	travelling wave chirp spectra,
since they are impossible from
	traditional sinusoidal spectra.
The chirps relate to
	sinusoidal wave spectra
as rotations over
	the local frequency-time planes at
		the source and the receiver,
the rotated frequency axes signifying
	phase accelerations,
		and equivalently clock accelerations.
The shifts result due to causality and
	the finite speed of light,
whose manifestation in the rotated view
resembles
	expansion of space.

The result finally reveals, and closes,
	a fine gap
between
	d'Alembert's general solutions
and
	Bernoulli's solution
		to the vibrating string problem
	as a series in sinusoidal waves
	\cite{Kleiner1989,Wheeler1987},
that has been thought complete
	because of Fourier theory,
but makes
	sinusoidal transport look
		fundamental and special.
The constancy of frequencies
	is often assumed
as sinusoidal wave solutions
	(\cf
	\cite[\S{1.3}]{BornWolf},
	\cite[\S{10-8}]{Goldstein}),
or obtained as eigenfunctions of
	time invariant Hamiltonians
	(\cf
	\cite[\S{10-3,4}]{Goldstein},
	\cite[\S{28,29}]{Dirac}).
However,
the stationarity of source dynamics
	or constancy of carrier frequencies
has no bearing on
	decomposition at a receiver,
which is strictly computational and dictates
	the spectral components seen,
and thus also lags
	in time varying component properties,
including
	frequency and wavelength in chirps,
which must be travel invariant
	to satisfy d'Alembert's equation.

The shifts then arise as
	chirp lags,
but empirical proofs
	were needed for
both
	the computational choice
and
	reality of the lags,
since
	the distance information is impossible
		per current theory.
The computational aspect
	and availability of
		distance information in waves
are specifically proved by
the \emph{absence} of the anomaly
	in the ESA Doppler analysis,
which uses
	a Fourier transform
	\cite{Jensen1998},
	in the Rosetta 2009 flyby
	\cite{Rosetta2009},
while its \emph{presence} in
	range data from the same signal,
demodulated using
	a carrier reconstructed with
		the Doppler rate,
required
	a false ephemeris correction
	\cite{Antreasian2015pvt}.

The SSN range residuals in the 1998 NEAR flyby and
	the ephemeris discrepancy in the Rosetta 2009 flyby
thus bear
	a fundamental significance
		complementing relativity,
of distinguishing
	a spectral reference frame
		from physical space-time.
The distinction decouples
	the wavelengths of reception or observation
from
	the source spectrum,
since the received spectrum can be
	arbitrarily shifted by
suitable choice of chirp frequency rates
	for any source distance,
so that
	tera-hertz or X-ray images can now be obtained
		under visible illumination,
	for example.
In communication,
	the capacity of a channel
is similarly defined by
	the sinusoidal assumption
	\cite{Shannon1949},
but signals of arbitrary wavelengths 
	could be received 
		simultaneously as chirp modes,
by using shifts to place them
	in the transmission band of
		the same optical fibre,
whose capacity would be then unlimited
	\cite{Prasad2005b,Prasad2008b}.

%stopzone }}}

% overview {{{

The SSN residuals and 
	their implication of excess delay
		in DSN and ESA data,
	are explained
		in the next section.
The theory of chirp travelling wave spectra
	is given next,
followed by
	quantitative analyses
of
	the SSN residuals
and
	the flyby anomaly,
substantiating
	the above.

%stopzone }}}

\section{Indication of the excess delay in DSN data} {{{

To an observer using
	an accelerating clock,
a sinusoidal wave should appear as
	a chirp having
		the reverse rate of change of frequency,
and chirps with
	the same frequency rate,
		as sinusoids.
Chirps waves necessarily exhibit
	frequency lags
that yield range in 
	continuous wave frequency modulated (CW-FM) radars.
In the accelerated clock view,
the chirp lags would appear as
	frequency shifts
which are impossible
	in sinusoidal waves,
and the shifts would be
	proportional to travel,
inconsistent with
	wave propagation as currently known.
Fig.~\ref{f:FlybyResiduals} is
	the graph of the SSN residuals
reproduced from
	\cite{Antreasian1998},
with scales of
	distance and one-way travel times for light
overlaid to expose
	their distance proportionality.

The
	$
	900
		~\metre
	$
	residual
at the start of tracking by the SSN
	is exactly the range error
that would occur in about
	$
	131
		~\milli\second
	$,
representing
	an optical path length of
	$
	33,000
		~\kilo\metre
	$,
at the radial speed of
	$
	6.870
		~\kilo\metre
		~\reciprocal{\second}
	$,
and it far exceeds
	the known resolutions of
	$15$-$25
		~\metre
	$
	at Altair
and
	$
	5
		~\metre
	$
	at Millstone
		SSN stations.
The negative sign is from
	the original graph
and can only mean that
	the NEAR spacecraft
		was that much closer,
according to SSN radars,
	than estimated by DSN.

The delays are also too large
	to blame radar processing.
Coherent radars perform
	phase correlated integration
only to extract
	weak echoes over noise.
The radar use of echoes for
	round trip timing
eliminates
	ambiguities of
		modulated range codes,
which get repeated and are periodic,
	but are the source of
		DSN and ESA range data.
The SSN datasets thus denote
	true round trip times,
and large errors solely during flybys
	would be in any case unlikely.
Occam's razor dictates,
	given the negative sign,
that the DSN signal had an excess delay
	impossible by current ideas,
but consistent with chirping
	due to acceleration,
as follows.

\FigNote{FlybyResiduals}\begin{figure}[h]
	\centering
	\psfig{file=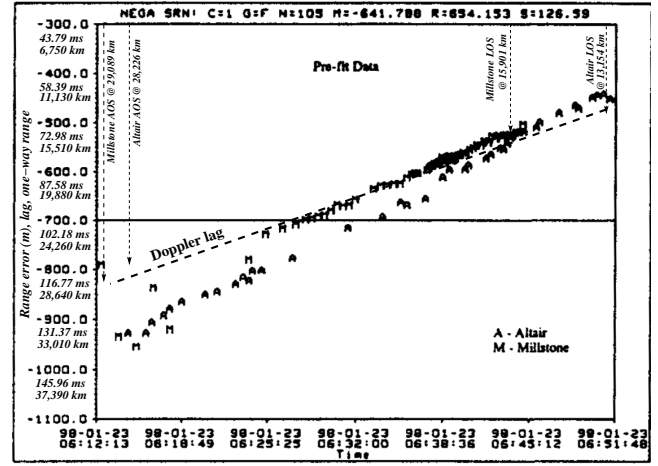, width=85mm}
	\caption{SSN residuals from \cite{Antreasian1998}
		with lag, range annotations}
	\label{f:FlybyResiduals}
\end{figure}

Denoting the instantaneous range errors
	as $\Delta r$,
and
	the radial speed
	as $v_o$,
the lag times in the figure
	are given by
	$
	\Delta t
	=
		\Delta r / v_o
	$,
and the one-way ranges,
	by
	$
	r
	=
		c \, \Delta t - r_e
	\approx
		c \, \Delta r / v_o - r_e
	$,
where
	$
	r_e
	\equiv
		6.371
		~\kilo\metre
	$,
	the earth's radius.
The slope of the residuals thus signifies
	proportionality of the range error to travel time
as
	$
	\Delta r
	=
		v \Delta t
	$.
The consistency of
	the DSN Doppler and differenced range data
	\cite{Antreasian1998,Anderson2008}
implies
	the same error affected the DSN Doppler.

The nonrelativistic two-way Doppler is
given by
	$
	\Delta \nu
	= 
		2 \nu v / c
	$
	at frequency $\nu$
for a velocity
	$v \equiv dr/dt$,
so
the DSN phase counters yielded
	smaller shifts
	$
	\Delta \nu'
	=
		2 \nu v' / c
	<
		\Delta \nu
	$.
The observed travel time proportionality
	more specifically
implies velocity error
	$
	d(\Delta r)/dt
	\equiv
		\Delta v
	=
		d(v \Delta t)/dt
	\equiv
		a \Delta t
	$,
where $a$ is
	the approach acceleration.
A Doppler lag can be only significant
	during a Doppler rate
	$
	d (\Delta \nu) / dt
	=
		2 \nu a / c
	$,
whose lag
	$
	d (\Delta \nu) / dt
		\times
		\Delta t
	$
would be therefore
	of frequency.

The uplink frequency was ramped to keep
	the downlink steady during the flyby
	\cite{Antreasian2015pvt},
so the delay and lags
	occurred in the uplink,
and were carried into the downlink by
	the phase-synchronous transponders onboard
	(\cf \cite[\S{III-A}]{Anderson2002a}).

The DSN carrier loop is designed to track
	the downlink carrier frequency continuously
even when its Doppler shift is changing
	(\cf \cite{DSNDoppler2010} \cite[\S{III}]{Anderson2002a}),
hence the DSN phase counts are of
	cycles of changing periods,
whereas Doppler theory was formulated for
	change in sinusoidal wave periods
	\cite{Doppler1842}.
The ESA extracts
	the Doppler
using a Fourier transform
	\cite{Jensen1998},
and thereby conforms to
	the sinusoidal definition
		even during accelerations,
since each output ``bin'' of
	a Fourier transform
is a count of cycles around
	a single frequency.
The bound of
	$
	4
		~\micro\metre
		~\reciprocal{\second}
	\pm
	44
		~\micro\metre
		~\reciprocal{\second}
		(1\sigma)
	$
stated against the anomaly
	in Rosetta's 2009 flyby
	\cite{Rosetta2009}
are just
	the resolution and phase noise
		in the ESA's Fourier transform.

The reconstructed carrier
	used for demodulation
		had to have been again a chirp,
	however,
		given the Doppler rate.
As Rosetta approached
	the earth along its orbital motion
		from behind
for gravitational boost
	(see \cite{Billvik2005} for
		all three flyby trajectory diagrams),
the earth would have receded
	over the excess delay in
		the range data.
The
	$
	13.34
		~\kilo\metre
		~\reciprocal{\second}
	$
	perigee velocity
and
	$
	2483
		~\kilo\metre
	$
	altitude
suggest
	$8
		~\milli\second$
	excess delay,
and
	$
	110.5
		~\metre
	$
	range error
as the magnitude of
	ESA's erroneous ephemeris correction.

In CW-FM radar,
the frequency lags yielding the range
comprise
	cumulative change of transmitter frequency
		over the radar pulse round trips.
Although the Doppler change
	was similarly continuous
in both
	pre- and post-encounter tracking segments,
and
	the modulated range codes yielded
		similar lags,
the reception process represents
	a maximum integration time $T$
shorter than
	a single bit in
		a modulated range code,
so the implied lags and frequency rate of
	the modulation side-band spectrum,
cannot have depended on
	integration through the round trip.
That is,
lags in a chirp spectrum
depend only on the instantaneous rate,
	and not a cumulative change of frequency,
		unlike cosmological shifts.

The residuals are thus evidence for
	chirp spectra
bearing lags exceeding
	the total carrier variation over
		the receiver integration times $T$,
and for realizability of
	fractional lags
	$
	z
	\equiv
		\Delta \nu / \nu
	\approx
		\beta r / c
	\gg
		\beta T
	$,
variation of 
	the receiver local oscillator (LO),
which followed the Doppler rate.

%stopzone }}}

\section{Chirp travelling wave spectra} % {{{

The general form of
	d'Alembertian solutions
	$
	f(r \pm ct)
	$
requires $f$ invariant of
	the retarded time
		$(t - r / c)$.
Invariance
	in $t$ or $r$ separately,
generally assumed for
	separating space and time parts
		of dynamical equations,
would be redundant for waves as
	the d'Alembertian solutions
		are already most general.
Rather,
as characteristic solutions defined by and for
	the constraint of constant frequencies,
sinusoidal waves
	were never most general.
The assumption of constancy
	avoided a problem,
however,
that any variation of frequencies with
	distance $r$ or time $t$
would make
	the received waves differ
from those
	observable at the source,
	\ie, at $t = r = 0$.

Yet,
any travel-invariant property $\xi$ of a travelling wave,
	hence other than amplitude or phase,
should be allowed to vary over time
	locally at points on the wave path,
and must then exhibit the lags
%\EqNote{distancelag}\begin{equation} \label{e:distancelag} %stopzone {{{
	$
	\Delta \xi
	\equiv
		\xi(t) - \xi(t - r / c)
	=
		\xi(t)
		- \dot{\xi} r / c
		+ \ddot{\xi} (r / c)^2 / 2!
		- \ldots
	\equiv
		\xi(t) [
			1
			- \beta r / c
			+ \beta^{(1)} (r/c)^2 / 2!
			- \ldots
			]
	$,
%\end{equation} %stopzone }}}
where
	$
	\beta
	\equiv
		\xi^{-1} d \xi / dt
	$,
	$
	\beta^{(1)}
	\equiv 
		\xi^{-1} d^2 \xi / dt^2
	$,
	$
	\beta^{(2)}
	\equiv 
		\xi^{-1} d^3 \xi / dt^3
	$
	\dots,
are fractional derivatives of $\xi$
	by the receiver's clock.
This is unlike the Hubble shifts,
	which are characterized using
		proper time along the path
	in current theory.

Fig.~\ref{f:WaveLags} shows that
such lags must occur in
	the wavelength of a chirp wave
because
	its local value around
		each crest and trough
	moves with the wave.
The lag
	$
	\Delta \lambda
	\equiv
		(\lambda_4 - \lambda_1)
	$
at time $t_2$
	at receiver $R$
must occur,
	in a locally measurable sense
		explained ahead,
as the waveform stays unchanged
	by travel.
The fractional shifts
	$
	z
	\approx
		1 - \beta r / c
	$
additionally imply
	time dilations,
via
	the Fourier inverse
\EqNote{timedilation}\begin{equation} \label{e:timedilation} %stopzone {{{
	\int_\Omega
		F(\omega [1 + z])
		\,
		e^{i \omega t}
	\, d \omega
%	\equiv
%		\int_\Omega
%			F(\omega')
%			\,
%			e^{i \omega' t / [1 + z]}
%		\,
%		%	d \omega'/[1 + z]
%		\frac{ d \omega'}{1 + z}
	=
		\frac{1}{1 + z}
%		(1 + z)^{-1}
		\,
		f
		\left(
			\frac{t}{1 + z}
%			t / [ 1 + z ]
		\right)
	,
\end{equation} %stopzone }}}
the amplitude factor denoting
	stretching of the energy
		over a dilated interval.
Equation (\ref{e:timedilation}) governs
	all uniform shifts,
including both
	Hubble shifts
and
	Doppler,
as highlighted recently by
	the Cassini-Huygens link failure
		as the signal dilation was overlooked
	\cite{Oberg2004}.
Dilations were not considered in
	Dirichlet's conditions,
which assured
	the completeness of Fourier theory
	\cite{Kleiner1989,Wheeler1987}.
As a receiver's local oscillators
	can be independently varied
		at arbitrary fractional rates $\beta$,
and would yield the corresponding chirp spectra
	as proved ahead,
the reconstructed waveforms
	would differ from the arriving waves
by arbitrary time dilations,
	which further depend on
		the distances of the individual sources!

%\FigNote{WaveLags}\begin{wrapfigure}[11]{r}{60mm}
\FigNote{WaveLags}\begin{figure}[h]
	\centering
	\psfig{file=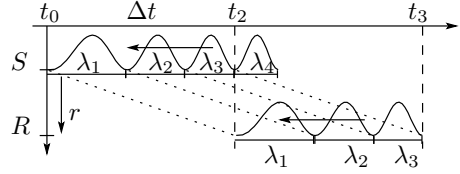, width=60mm}
	\caption{Lags in chirp waves}
	\label{f:WaveLags}
\end{figure}

As a prediction in a differential form
	from radar imaging
	\cite{Prasad2005b,Prasad2008b},
this had made no sense
	and seemed causally flawed
	\cite{Elmegreen2009pvt}.
It is finally explained by 
	the computational character of
		a chirp spectrum
	in Fig.~\ref{f:FreqLags},
as a rotation of
	the receiver's local
	frequency
		($RR_\Omega$)
	and
	time
		($-RR_T$)
	axes,
denoting
	the local evolution of
		the spectral components
			in time
	by the receiver's clock.
The constant frequency of
	a sinusoidal component
would be represented by
	vertical lines
		like $BC$.
The inclined lines
	$GC$, $HF$
denote
	chirp components
with frequencies
	increasing over time.
The spectrum at present time $t_2$
is represented by
	the same coefficient values on
		the frequency axis $RR_\Omega$
regardless of
	the inclination.

With the inclination, however,
excess one-way delays are incurred,
	just as in the DSN Doppler,
that result in shifts exactly equal to
	cumulative change from
		an earlier state at the source,
so the distance information
	bears the penalty of excess delay.
Each chirp line,
	projected indefinitely,
not only attains
	every possible frequency
		at some instant,
but is identical to
	every other chirp of the same inclination
by a simple displacement
	in time.
This equivalence leads to
	the excess delay,
as the travel delay acts against
	the frequency change.
Conversely,
were the angle of inclination
	$
	\angle CBF
	\equiv
		\tan^{-1}
		(
		|CF| / |BC|
		)
	=
		\tan^{-1}
		(
		\beta \Delta t
		/
		\Delta t
		)
	=
		\tan^{-1}
		\beta
	$
	made $0$,
the chirps would become
	degenerate vertical lines
		through $C$ and $F$
that overlap no longer
	if displaced in time,
so the delay and the distance information
	both vanish.

These details,
	and relations to causality
		and the speed of light,
are revealed by incorporating
	the source
		frequency ($SS_\Omega$)
	and
		time ($-SS_T$)
	axes,
with
	corresponding source chirp lines
		$JD$ and $AE$
	parallel to
		$GC$ and $HF$.
Sinusoidal transport
	would be represented by
parallel lines like $DC$ and $EF$
	connecting equal values on
		the source and receiver frequency axes.
Hubble's law would require
	inclined lines like $EC$
to produce shifts
	$
	\Delta \omega
	=
		\omega_4 - \omega_1
	=
		|CF|
	$
	at distance $r$
and
	$
	\Delta \omega_1
	\equiv
		|LM|
	\approx
		|CF| r_1 / r
	$
	at distance $r_1 \equiv |EM|$.

%\FigNote{FreqLags}\begin{wrapfigure}[18]{l}{65mm}
\FigNote{FreqLags}\begin{figure}[h]
	\centering
	\psfig{file=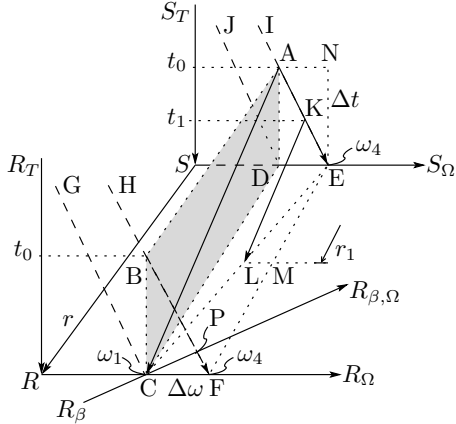, width=60mm}
	\caption{Spatial relation of spectral histories}
	\label{f:FreqLags}
\end{figure}

The $r - ct$ invariance required of
	d'Alembertian solutions
more particularly requires
	lines like $AC$  and $KL$
inclined at
	$
	\angle DAC
	=
		\tan^{-1}( |DC|/|AD| )
	=
		\tan^{-1}( r / \Delta t )
	\equiv
		\tan^{-1} c
	$
with respect to
	the time axes,
but parallel to
	the distance vector $SR$,
so as to connect equal component frequencies of
	the source current and receiver voltage spectra,
regardless of whether
	the connected frequencies belong
to chirps,
	as denoted by lines $IE$ and $HF$,
or to sinusoids,
	represented by lines $AD$ and $BC$,
		respectively.
The inclination denotes
	wave speeds $c < \infty$,
and is along of increasing time
	from source ($A$) to receiver ($C$),
		conforming to causality.

More importantly,
a component with angular frequency $\omega_1$ at $C$
	on chirp line $GC$
should correspond to
	the same angular frequency $\omega_1$ in
		source history ($A$),
but belong on
	chirp line $IE$
that changes to $\omega_4$
	by time $t_2$ ($E$).
However,
an atom emitting at
	angular frequency $\omega_4$ at $t_2$ ($E$)
would have been observed locally
	at $\omega_4$ also at $t_0$ ($N$),
and the same should hold for
	a steady carrier transmission.
It thus appears that
	the d'Alembertian travel lines
		like $AC$
either require amplitudes to shift with travel,
	from $N$ to $C$,
which would conflict with
	the d'Alembertian invariance;
or chirp spectral decompositions,
	which can only produce
		inclined histories like $HF$ and $AE$,
	must be impossible,
so the lags $\Delta \omega$ would require
	nonlocal simultaneous measurements
		at source and receiver
	at $t_2$.
The second case is
	cannot hold
since the inclinations $\beta$
	could be infinitesimally small,
and the decomposition is
	in any case purely computational.

The answer is that the construction
already implies that
	at nonzero $\beta$,
$\omega_1$ is seen
	\emph{only} at distances
	$
	r
	= 
		c \, \Delta \omega / \beta
	$.
The amplitude at $C$ comes from $A$,
	which is precursor
		to $E$ at $t_0$
	and
		to $N$ at $t_0 - \Delta t$.
The chirp spectrum thus reconstructs
	distributions at the past times
	$
	t_0 - \Delta t
	\equiv
		t_2 - 2 \Delta t
	$,
where the factor $2$ relates to
	the excess delay.

Chirp spectra would be thus
	time invariant like Fourier spectra,
but exhibit
	distance proportional shift factors and dilations
with the receiver's choice of
	$\beta$ and its derivatives,
\emph{because the chirp spectra start
	fully shifted and dilated at the source!}
The total energy is also
	clearly unchanged.

The inclined axis $R_\beta R_{\beta,\Omega}$,
denotes
	the chirped spectral view,
given by
	the DSN and ESA range data during flybys,
in which
	local chirp histories $GC$ and $HF$
		seem normal to the frequency axis,
but travel lines $AC$, $KL$
	unaccountably seem inclined.
The inclination of axis is equivalent to
	the receiver's clock acceleration
		inclining the components;
the segment $|PF$ denotes
	the relative phase accelerations $|PF|$
that are not apparent in
	the rotated ``reference frame'',
in which
	the chirps appear as
a Fourier spectrum with shifts
	$|CP| \approx \Delta \omega$,
due to
	skewing of all travel lines $AC$, $KL$
to longer wavelengths,
	as if space itself were expanding.

%stopzone }}}

\section{Reception and orthogonality} % {{{

In any frequency modulation scheme,
including
	phase shift keying (PSK)
		in deep space telemetry
	\cite{DSNTelemetry2009},
can be described by
	a random variable 
	$\Omega_m$
denoting
	the instantaneous modulation.
Both at the DSN receiver
	and the spacecraft transponder,
the carrier loop phase locks imply,
	upon allowing for frequency variations,
the first order product integral condition
\EqNote{modchirporth}\begin{equation} \label{e:modchirporth} %stopzone {{{
\begin{split}
	%\frac{1}{ 2 \pi }
	\int_{T}
	%\Exp{
		\exp\left[
			\frac{i}{\beta}
			( \omega_c + \Omega_m )
			%\beta'^{-1}
			e^{ \beta' ( t - r / c ) }
		\right]
		\exp\left[
			- \frac{i \omega_o}{\beta}
			e^{ \beta t }
		\right]
	%}
	dt
\\
	\simeq
	2 \pi
	\,
		\delta\left(
			\omega_c + \Exp{ \Omega_m }
			- 
			\omega_o
		\right)
		\delta(
			\beta'
			-
			\beta
		)
	,
\end{split}
\end{equation} %stopzone }}}
where
	$\omega_c$ is
		the carrier;
	$\omega_o$ is
		the loop voltage-controlled oscillator (VCO) frequency;
	$\beta$ and $\beta'$ are
		fractional rates of the VCO and
				a received spectral component,
			respectively;
and
	$T$ is the loop filter time constant.
$T$ is set
	below $1~\hertz$ in DSN carrier loops
in order to suppress both
	phase noise and modulation
	\cite{DSNDoppler2010}.
The $\beta^{-1}$ factor
	is from integrating
the exponential chirp
	$
	\omega(t)
	=
		\omega_0
		e^{\beta t}
	$
	for get the phase,
and vanishes
	in the phase derivative
		via L'H\^opital's rule.

Equation (\ref{e:modchirporth}) constitutes
	the orthogonality condition for
		exponential chirp waves
without modulation
	($\Omega_m = 0$),
and including the case of
	$T \rightarrow \infty$,
since a travelling wave of the same
	instantaneous frequency and rate of change
as
	the receiver's LO ($\sim \omega_o$)
contributes in every cycle to
	the integration performed by
		subsequent filters,
but any other component contributes
	over at most a cycle.
In a Fourier transform,
nonmatching components contribute
	at every few cycles indefinitely,
so Fourier convergence
	depends on Ces\`aro means,
		and is weaker in this sense.

The orthogonality looks weak
	for distinguishing between
say,
a chirp of fractional rate $1~\reciprocal{\second}$
	at $5~\tera\hertz$
from
	a $5~\tera\hertz$ sinusoid
as their phases would differ by only
	$10^{-7}~\radian$
over
	$10^5$ cycles,
but in
	a spectral selection or decomposition,
all families of curves
	over local frequency-time planes (Fig.~\ref{f:FreqLags})
		must be assumed available.
The contributions from
	$\beta \pm \delta \beta$ pairs
then cancel out
	for $\delta \beta \ne 0$,
just as in
	the interference of alternative paths
		in Fermat's principle.
For decoding or demodulation,
equation (\ref{e:modchirporth}) relates
	the modulated carrier and LO statistically
over shorter integration times $T$
	for the modulation bandwidth,
assuming
	$\Exp{\Omega_m} = 0$,
since 
	the \dc{}(direct current) is suppressed
		in deep space telemetry.
The consistency of
	the DSN range data with its Doppler
implies that
	the modulated chirp spectrum had
		the correct phase offsets $\Omega_m$
	relative to the chirp carrier.

Larger lags, of
	$9~\mega\hertz$
at
	$1~\AU$
for the same acceleration
	$
	0.5
		~\metre
		~\second^{-2}
	$
as at loss of signal (LOS), would shift
	the chirps out of the filter pass-bands,
so the signal presumably gets demodulated from
	the Fourier spectrum
		without lags%
	\footnote { % {{{
	The lags should be about $18~\hertz$
		at lunar range,
	but inband chirps would then face 
		echo suppression
	due to
		their $1.2~\second$ excess delay.
	A carrier loop lock to
		the Fourier spectrum
	would produce
		a piecewise frequency approximation of
			the Doppler rate
		as the VCO carrier,
	so each range code bit then gets retrieved
		from the Fourier spectrum.
	}. %stopzone }}}

%stopzone }}}

\section{Explanation of the SSN residuals} % {{{

The net gain in speed was only
	$
	(6.87 - 6.83) / 6.87
	\approx
		0.6\%
	$
	\cite[Fig.~3a]{Anderson2008},
with most of the acceleration
	close to earth
after the SSN tracking
	in increasingly tangential motion.
The uniformity of
	the $10~\minute$ ticks
	in the equatorial view
		\cite[Fig.~1]{Anderson2008}
and of similar ticks in
	the north polar view
		\cite[Fig.~9]{Antreasian1998},
which are expanded due to 
	projection,
suggest that
	the mean speed
	$
	v_o
	\equiv
		6.85 
		~\kilo\metre
		~\reciprocal{\second}
	$
would be adequate for
	present purposes.

The $219~\minute$ gap in the DSN tracking
	then represents
	$
	6.851 
		~\kilo\metre
		~\reciprocal{\second}
	\times
		219
		~\minute
	=
		90,000
		~\kilo\metre
	$
	of trajectory.
LOS occurred
	$
		1\hour
		~
		8\minute
	$
	before periapsis
and acquisition of signal (AOS),
	at Canberra,
at
	$
		2\hour
		~
		31\minute
	$
	after periapsis,
so
the range was
	$
	90,000
		~\kilo\metre
	\times
		68 / 219
	\approx
		27,950
		~\kilo\metre
	$
	at LOS
and
	$
		62,070
		~\kilo\metre
	$
	at AOS.
Tracking at Altair ended
	$
		36~\minute
	$
past LOS at 06:51:08
	and had started at 06:14:28,
for a total of
	$2200~\second$,
so the tracking started
	$
	4120~\second
	$
	before periapsis,
at
	$
	r
	\approx
		4120
	\times
		6.851 
		~\kilo\metre
		~\reciprocal{\second}
	=
		28,226
		~\kilo\metre
	$.
The one-way delay was therefore
	$
	\Delta t
	\equiv
		- r / c
	\approx
		- 94
		~\milli\second
	$,
implying
	range error
	$
	\epsilon_r
	\equiv
		v \, \Delta t
	=
		-
		94
			~\milli\second
		\times
		6.851 
			~\kilo\metre
			~\reciprocal{\second}
	\approx
		-
		645
		~\metre
	$,
about $25\%$ smaller than
	in Fig.~\ref{f:FlybyResiduals}.
The error decreased with
	the range rate
at
	$
	d \epsilon_r / dt
	\approx
		v \, d (\Delta t) / dt
	=
		v^2 /c
	=
		(6.851 
			~\kilo\metre
			~\reciprocal{\second})^2
		/ c
	\approx
		0.313
		~\metre
		~\reciprocal{\second}
	$,
over 
	$
	1187
		~\second
	$
from
	06:25:25
to
	06:45:12,
hence by
	$
	0.313
	\times
		1187
	\approx
		186
		~\metre
	$,
which is within $10\%$ of the 
	$
	200
		~\metre
	$
	Millstone decrease.
Fig.~\ref{f:FlybyResiduals} shows
	two sets of residuals,
since they are projections of
	the same lag of
		the (pre-LOS based) DSN estimate
	behind
		the true trajectory
in the direction of
	each of the two SSN stations.
The ground track diagram
	\cite[Fig.~7]{Antreasian1998}
shows the trajectory pointed towards Millstone
	initially,
implying
	a faster initial decrease of range,
hence
	greater initial values
		for Millstone,
	as seen in Fig.~\ref{f:FlybyResiduals}.

%stopzone }}}

\section{Explanation of the flyby anomaly} % {{{

The delay means that
DSN underestimates
	pre-encounter approach speed
and overestimates
	post-encounter recession
and thus infers
	an anomalous velocity gain in
		earth flybys
whenever the tracking is discontinuous
	across periapsis.
If tracked continuously,
however,
	the delay in the Doppler's change of sign
		at periapsis
	(Fig.~\ref{f:FlybyLags})
should cause
	a negative $\Delta v$.

The negative $\Delta v$
	in Galileo's second flyby
was concluded from
	around periapsis,
since it was at first thought
	masked by atmospheric drag
	\cite{Antreasian1998,Anderson2008}.
The tracking was unbroken
	in Cassini's flyby
that also showed negative $\Delta v$
	\cite{Burton2001}.

As the excess delay
	varies with range,
the true velocity profile,
	given by
		the differenced SSN range,
	and
		the DSN Doppler
would be closest at periapsis,
	and cannot really be parallel.
The slopes of the residuals were thus
	``irreducible through velocity estimation''
	\cite{Antreasian1998},
though both curves were monotonic
	over the SSN tracking period,
as shown.

%\FigNote{FlybyLags}\begin{wrapfigure}[10]{l}{80mm}
\FigNote{FlybyLags}\begin{figure}[h]
	\centering
	\psfig{file=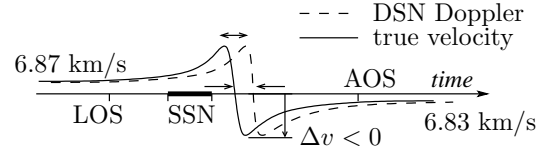, width=70mm}
	\caption{DSN Doppler and its lags during flybys}
	\label{f:FlybyLags}
\end{figure}

The velocity error at AOS should cause
	post-encounter data
to be inconsistent with
	the pre-encounter trajectory,
		and \viceversa.
Acceleration due to earth's gravity
	at AOS range
would be
	$
	a
	=
		0.103
		~\metre
		~\second^{-2}
	$,
implying
	a velocity error
	$
	\Delta v
	=
		- a \Delta t
	\approx
		21.4
		~\milli\metre
		~\reciprocal{\second}
	$,
or
	a
	$
		603
		~\milli\hertz
	\sim
		10.7
		~\milli\metre
		~\reciprocal{\second}
	$
Doppler amplitude
	at the downlink frequency.
These are about $20\%$
	of the reported
	$
		760
		~\milli\hertz
	=
		13.5
		~\milli\metre
		~\reciprocal{\second}
	$
	\cite{Anderson2008}.

Canberra's latitude of
	$35.2828\degree$
means it is
	$
	6371~\kilo\metre
	\times
		\cos(
		35.2828\degree
		)
	=
		5201
		~\kilo\metre
	$
	off the earth's axis.
The $-71.96\degree$ declination of
	the post-encounter velocity asymptote
then implies
	$
	5201
		~\kilo\metre
	\times
		\cos(
		71.96\degree
		)
	\approx
		1611
		~\kilo\metre
	$
	of diurnal range
and
	$
	(1611/62070) \times 603
		~\milli\hertz
	\approx
		15.6
		~\milli\hertz
	$
	diurnal Doppler oscillations.
The larger actual $50~\milli\hertz$ amplitude
is due to
	a smaller declination at AOS,
and to a misprediction of direction 
	\cite{Anderson2008},
possibly worsened by
	the error at LOS%
	\footnote { % {{{
	$
	a
	\approx
		0.51
		~\metre
		~\second^{-2}
	,
	\Delta t
	\approx
		93
		~\milli\second
\Rightarrow
	\Delta \nu
	\approx
		47.6
		~\milli\metre
		~\reciprocal{\second}
	\equiv
		1.34
		~\hertz
	$.
	}. %stopzone }}}

The velocity error
	$
	\Delta v
	=
		- a \Delta t
	\equiv
		- a r / c
	$
also explains
	the $r^{-1}$ decay
in the post-encounter oscillation graphs
	\cite{Antreasian1998,Anderson2008},
since $a \propto r^{-2}$
	due to earth's gravity.

%stopzone }}}

\section{Conclusion}% {{{

All features of the flyby anomaly 
	are thus explained by
a delay proportional to range
	in Doppler and range data
derived from
	the telemetry signal
that was chirped
	due to the Doppler rate,
and should be impossible by
	current notions of wave propagation.

More particularly,
the distance proportionality of 
	the delay
and of
	the equivalent frequency lags
		in the telemetry spectrum
are given by
	two independent radar tracks
which had been overlooked for
	over a decade,
oddly,
in the very quest for deviations
	from the known laws.

The chirping and lags
	should be also easy to verify
over ground distances
	at radio frequencies,
with no motion
	or the difficulties of
		optical implementation
	(\cf \cite{Prasad2008b}).

%stopzone }}}

% ------------------------------------------------------------------
\begin{raggedright}

\end{raggedright}
\end{document}